\documentclass[preprint2]{aastex62}

\usepackage{epsfig}
\usepackage{natbib}
\usepackage{amsmath}

\newcommand{\kms}{km~s$^{-1}$}
\newcommand{\Msun}{$M_{\odot}$}

\newcommand{\kmsMpc}{km~s$^{-1}$~Mpc$^{-1}$}

\begin{document}
\title{Cosmicflows-3: Two Distance$-$Velocity Calculators}

\author{Ehsan Kourkchi}
\affil{Institute for Astronomy, University of Hawaii, 2680 Woodlawn Drive, Honolulu, HI 96822, USA}
\author{H\'el\`ene M. Courtois}
\affil{University of Lyon, UCB Lyon 1, CNRS/IN2P3, IUF, IP2I Lyon 69622, France}

\author{Romain Graziani}
\affil{University of Lyon, UCB Lyon 1, CNRS/IN2P3, IUF, IP2I Lyon 69622, France}

\author{Yehuda Hoffman}
\affil{Racah Institute of Physics, Hebrew University, Jerusalem, 91904 Israel}
\author{Daniel Pomar\`ede}
\affil{Institut de Recherche sur les Lois Fondamentales de l'Univers, CEA, Universite' Paris-Saclay, 91191 Gif-sur-Yvette, France}
\author{Edward J. Shaya}
\affil{University of Maryland, Astronomy Department, College Park, MD 20743, USA}
\author{R. Brent Tully,}
\affil{Institute for Astronomy, University of Hawaii, 2680 Woodlawn Drive, Honolulu, HI 96822, USA}

\begin{abstract}
Tools are provided at the Extragalactic Distance Database website that provide relationships between the distances and velocities of galaxies based on smoothed versions of the velocity fields derived by the Cosmicflows program.
\end{abstract}

\bigskip

\smallskip
\section{Introduction}
Galaxy velocities deviate from Hubble-Lema\^itre expansion.  Deviations can be considerable, as evidenced by the motion of the Local Group of 631~\kms\ with respect to the rest frame of the cosmic microwave background \citep{1996ApJ...473..576F}.  There are numerous instances, particularly nearby, when it is useful to have a better approximation between observed velocities and physical distances than provided by the simple assumption of uniform cosmic expansion.

The NASA/IPAC Extragalactic Database (NED) has provided estimates of galaxy distances given observed velocities based on a model by \citet{2000ApJ...529..786M}.  In alternatives, deviations are induced by up to three mass concentrations, associated with the Virgo Cluster, the Great Attractor, and the Shapley Concentration.  The parameters of this model are the positions of the mass centers, the velocities induced at our location by each, and the assumption that the masses have spherically symmetric geometry with density gradients $\rho(x) \propto r^{-2}$ where $r$ is the distance from a mass center.
 
This model has a direct lineage from models by \citet{1988lsmu.book..115F}, \citet{1990ApJ...360..448H}, and \citet{1992ApJ...395...75H}. The latter of these, although it does not entertain the Shapley Concentration, considers added details, the most important being the Local Velocity Anomaly \citep{1988Natur.334..209T, 1988lsmu.book..115F}.  These early studies surmised that this feature is related to the proximity of the Local Void  (see also \citet{1988MNRAS.234..677L}), a proposal that has now been robustly confirmed \citep{2017ApJ...835...78R, 2019ApJ...880...24T,2019ApJ...880...52A}.  Already, then, the \citet{1992ApJ...395...75H} model and variants capture the major features affecting the local velocity field, although another player to have emerged is the vast underdensity at the cosmic microwave background dipole anti-apex; the Dipole Repeller \citep{2017NatAs...1E..36H} and the Perseus$-$Pisces filament \citep{1988lsmu.book...31H} significantly inhibits  the flow in the CMB dipole direction.

Many other contributions could be entertained (Coma, Horologium-Reticulum, Hercules, ...).
It should be clear that any parametric model with a moderate number of parameters will only crudely approximate the peculiar velocity field.  Also, for a model to be useful to the community, it should be relatively painless and efficient for a user to acquire desired information for a random target.  The facility to be described is based on a velocity field responding to the full complexity of structure on scales $1 - 200$~Mpc.  Two models are offered.  One restricted to 38~Mpc is based on a fully non-linear analysis.  The other extending to 200~Mpc is derived from an analysis in the linear dynamical regime.  Both are publicly accessible at the interactive platform to be described.

\section{Distance$-$Velocity Users Manual}

Details of the two underlying models will be discussed in the next section, but the user interface functions are the same for both.  The facility can be accessed at the Extragalactic Distance Database (EDD).\footnote{\url{http://edd.ifa.hawaii.edu}} 
\footnote{Pre-publication, access to the linear model calculator is at
 \url{http://edd.ifa.hawaii.edu/CF3calculator}}

\begin{figure*}[]
\centering
\includegraphics[width=0.70\linewidth]{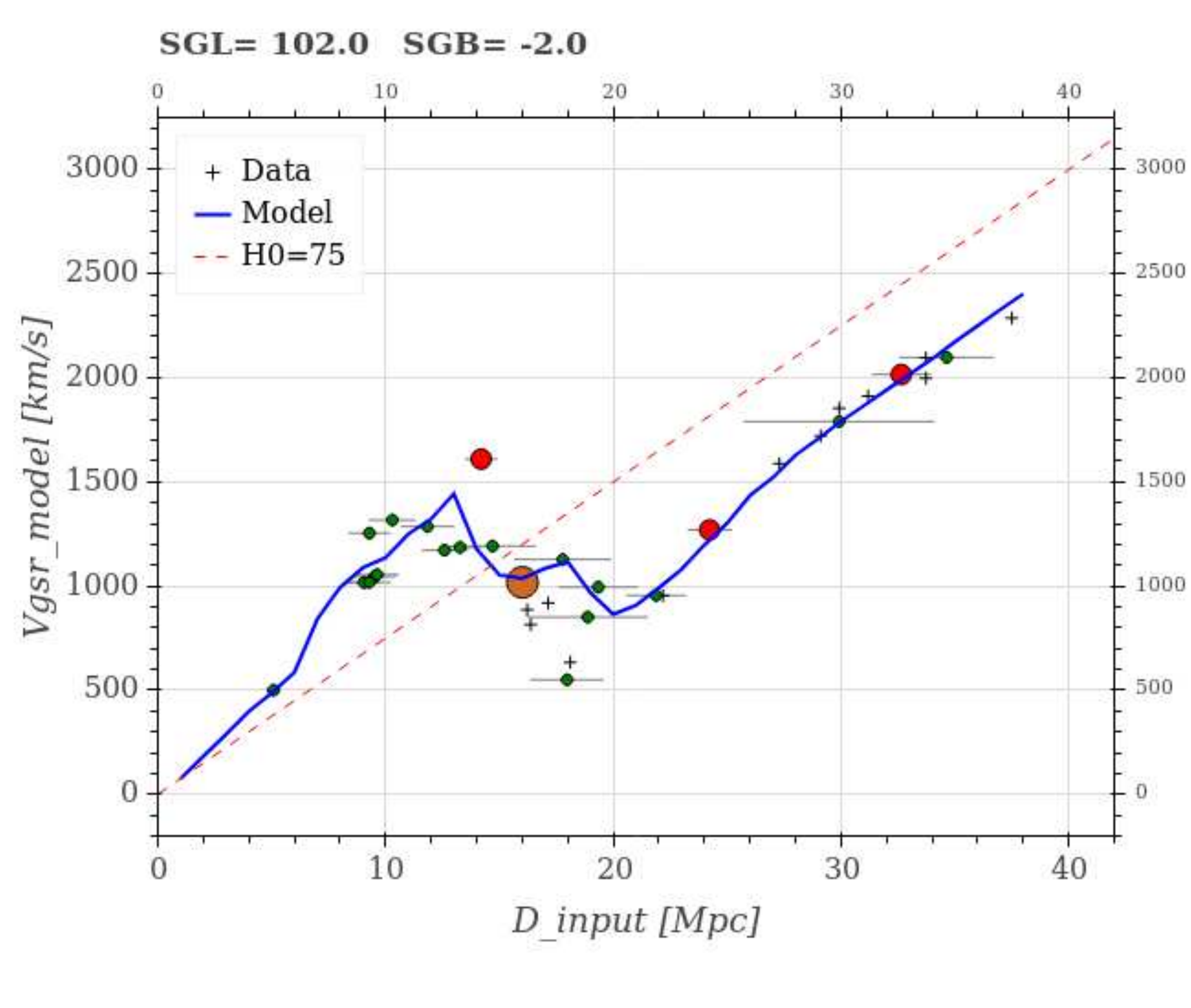}
\caption{
Velocities as a function of distance in the direction of the Virgo Cluster plotted with the non-linear numerical action calculator.   The Virgo Cluster lies at 16.0~Mpc with $V_{gsr} = 1096$~\kms.  The blue curve from the model is noisy because the number of local constraints is small at each step in the averaging but the characteristic triple-value curve (3 distances sharing the same observed velocity) associated with infall around a massive cluster is clear.   The red dashed line assumes H$_0=75$~\kmsMpc.  Data constraining the model are shown as large red circles if distance uncertainties are 5\% or less (brown extra large circles for Virgo and Fornax clusters), and as small green circles if uncertainties are over 5\% but not more than 15\%. Data with larger uncertainties are represented by black plus signs but do not constrain the model.
}
\label{DVnam}
\end{figure*}

\begin{figure*}[]
\centering
\includegraphics[width=0.65\linewidth]{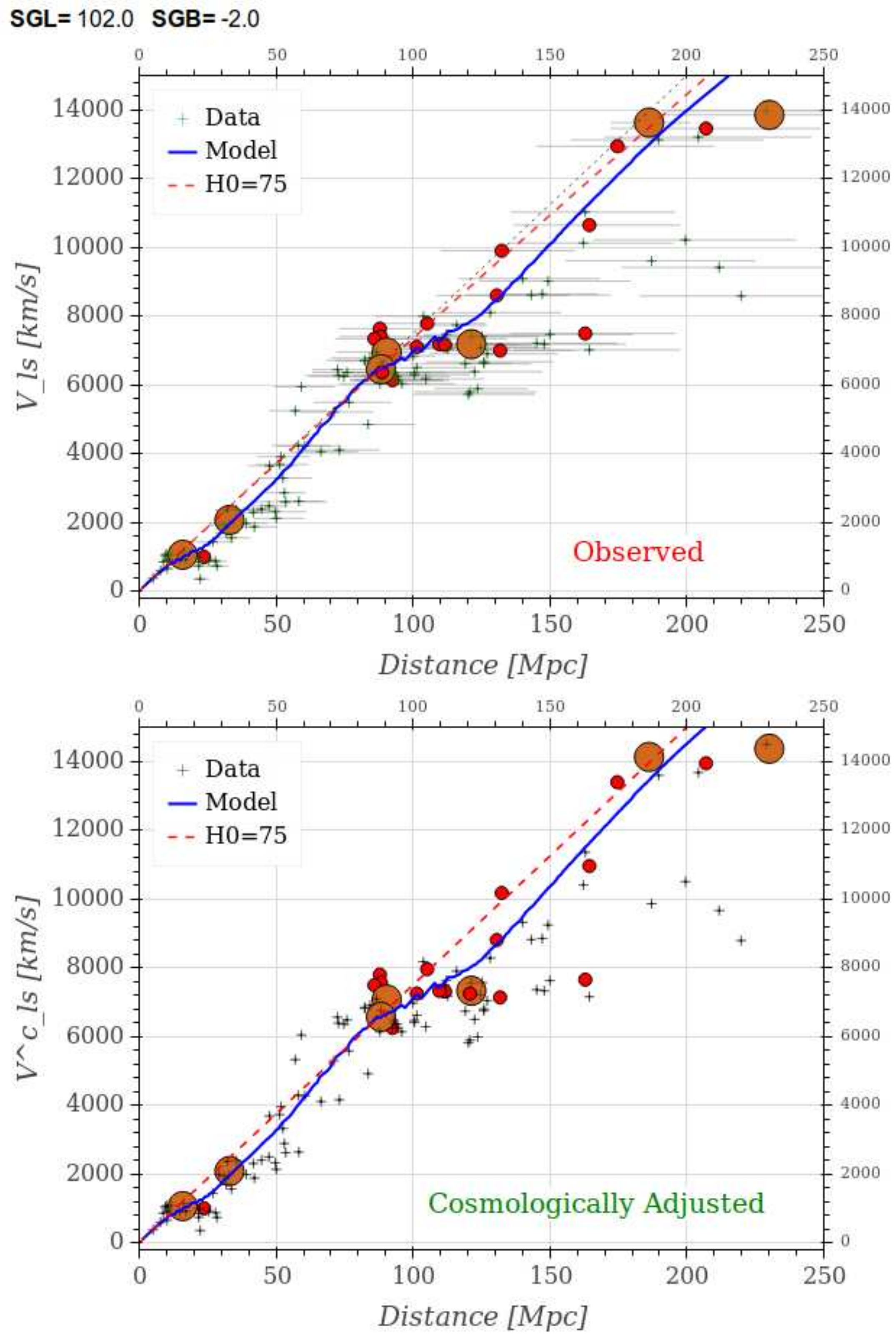}
\caption{
Velocities as a function of distance plotted with the linear calculator. Cosmological corrections become noticeable in this extended domain; the two panels give alternate representations (see text). The blue curve is derived from interpolation of velocities on a grid of distances with intervals of 6.25 Mpc, where the value at each point is averaged over the 8 nearest grid points weighted by the inverse square of separation. In the top panel, velocities are the observed values in the Local Sheet frame and the red dashed locus of constant H$_0$ has curvature away from the grey dotted straight line.  In the bottom panel, velocities are adjusted for the cosmological effect and the red dashed Hubble line is straight. In the two panels of this figure, the data are coded by the mass of entities: large brown circles if mass greater than $10^{14}$\Msun, smaller red circles if mass is between $10^{13}$ and $10^{14}$\Msun, and black plus signs if mass is below $10^{13}$\Msun.  The coordinates are chosen to be the same is in Fig.~\ref{DVnam}. Major clusters in the cone in this direction are the Virgo Cluster at 16 Mpc and Abell 1367 at 88 Mpc.
}
\label{DVlin}
\end{figure*}

A user can enter the celestial coordinates of a target with a choice of coordinate systems.  
Examples are shown in Figures~\ref{DVnam} and \ref{DVlin}.
Observational data from the {\it Cosmicflows-3} compendium of distances \citep{2016AJ....152...50T}  can be displayed in a cone chosen by the user centered on the target.  A blue locus plots the averaged expectation velocity along the line of sight as a function of distance.\footnote{All distances in this tool are luminosity distances, $d_L$, related through redshift, $z$, to comoving distances, $d_m$ by the equation $d_L = d_m (1+z)$.} Cursor control can access specific distance and velocity values along the locus.  Hovering over a datum gives name, velocity, and distance specifics for that item.  Zoom and translation functions are activated by side-panel toggles.

The red dashed straight line in every plot shows the relationship between velocity and distance with uniform expansion if the Hubble Constant is 75~\kmsMpc. This line is for illustrative purposes only!  Velocities and distances are determined completely independently of each other in the {\it Cosmicflows-3} compilation.  The models make no assumption about the value of H$_0$.

The distance$-$velocity tool does {\bf not} directly give peculiar velocities, $V_{pec}$. The tool gives expectation distances, $d$, at observed velocities, $V_{obs}$ (or expectation observed velocities at specified distances). To a reasonable approximation, $V_{obs} = {\rm H}_0 d +V_{pec}$.\footnote{\citet{2014MNRAS.442.1117D} discuss the more rigorous formulation $V_{pec}=(f(z) V_{obs} - {\rm H}_0 d) / (1+ {\rm H}_0 d /c)$ where $f(z)$ is given by Eq.~2.  Differences from the approximate formula grow with $z$ to $\sim 30$~\kms\ by $z=0.05$.}  It has been demonstrated \citep{2016AJ....152...50T} that {\it Cosmicflows-3} distances and velocities are compatible with H$_0 = 75$~\kmsMpc, with zero-point uncertainty at the level of 3\%.  Users interested in peculiar velocities can contemplate applying alternative values of H$_0$ at their risk.  Note that altering the zero point associated with {\it Cosmicflows} distances alters the H$_0$ value that would be most consistent but the products H$_0d$ would be unchanged.  Hence, the relationship between $V_{obs}$ and $V_{pec}$ given in the formula above is independent of the zero point calibration.

The reference velocities are different for the two models.  In the case of the local non-linear model, velocities are with respect to the Galactic center.  With the large scale linear model, velocities are with respect to the Local Group.  Justifications for these choices will be given in the section discussing the models.  In any event, it must be noted that there are variants of both these reference frames in the literature.  The definition of the Galactic Standard of Rest (gsr) depends on the distance of the Galactic center and the amplitude of Galactic rotation at the Sun, as well as a small solar deviation.  There are alternative choices \citep{1991trcb.book.....D, 2005ApJ...628..246E, 2009ApJ...700..137R}.  Our model is based on the solution by \citet{2012ApJ...753....8V}.  As a measure of the uncertainties inherent in the translation to the Galactic rest frame, differences between the four cited alternatives are $12\pm5$~\kms.

Similarly, there are alternative versions of the Local Group rest frame \citep{1977ApJ...217..903Y, 1996AJ....111..794K, 1999AJ....118..337C}.  We prefer a variant we call the Local Sheet (ls) reference frame \citep{2008ApJ...676..184T}.  The three former (Local Group) solutions are derived from the properties of galaxies within 1~Mpc.  The Local Sheet solution is derived from the properties of galaxies {\it outside} 1~Mpc yet within 5~Mpc.  See \citet{2008ApJ...676..184T} for the argument that this solution is the most stable.  Uncertainties between the alternative Local Group frames are at the level of $19\pm10$~\kms.

The linear model extends to velocities of 15,000~\kms\ and cosmological corrections reach $\sim4\%$.  The corrected velocity $V_{ls}^c$ is related to the observed velocity $V_{ls}$ by:
\begin{equation}
    V_{ls}^c = f(z)V_{ls} 
\end{equation}
where 
\begin{equation}
    f(z) = 1+1/2(1-q_0)z-1/6(2-q_0-3q_0^2)z^2
\end{equation}
\begin{equation}
    q_0 = 1/2(\Omega_m-2\Omega_{\Lambda}) = -0.595 
\end{equation}
when $\Omega_m=0.27$ and $\Omega_{\Lambda}=0.73$ and $z=V_{ls}/c$ \citep{2006PASP..118.1711W}.  

In the top panel of Figure~\ref{DVlin} velocities are observed values (averaged over group members) but in this representation there is a slight curvature of the Hubble relation manifested in the departure of the red dashed curve from the faint grey straight line.  Velocities in the bottom panel are increased by the cosmological correction so the red dashed Hubble line is straightened.  In detail the correction depends on the choices of the cosmological matter and energy density parameters $\Omega_m$ and $\Omega_{\Lambda}$ but differences in the correction between reasonable values of these parameters are negligible within the velocity range being considered. $f(z=0.05)$ decreases by $0.1\%$ if $\Omega_m$ increases and $\Omega_{\Lambda}$ decreases by 0.03.

Hover with the cursor over a datum element of the linear model to obtain input distance and velocity information.  The element may be an individual galaxy or a group and is identified by the parameter PGC1, the Principal Galaxies Catalog identification \citep{1996pgcs.book.....P} of the brightest member. The Nest parameter identifies the membership of elements in the 2MASS group catalog of \citet{2015AJ....149..171T}.

For convenience, formulae are given here for conversions between heliocentric velocities, $V_h$, and the reference frames used in the distance$-$velocity tools where the angles are Galactic longitude and latitude ($l,b$).

\noindent
Galactic standard of rest:
\begin{equation}
V_{gsr} = V_h+11.1\cos{l}\cos{b}+251\sin{l}\cos{b}+7.25\sin{b}
\end{equation}
Local Sheet:
\begin{equation}
V_{ls} = V_h-26\cos{l}\cos{b}+317\sin{l}\cos{b}-8\sin{b}
\end{equation}
Between Galactic standard of rest and Local Sheet:
\begin{equation}
V_{ls}=V_{gsr}-37\cos{l}\cos{b}+66\sin{l}\cos{b}-15\sin{b}
\end{equation}

\section{Models}

Two calculators are offered; one limited to distances less than 38~Mpc ($V_{obs} \sim 2850$~\kms) based on a fully non-linear dynamical model and the other extending to 200~Mpc (15,000~\kms) at lower resolution and based on a linear model.

\subsection{Non-linear Model Limited to 38~Mpc}

This calculator presents a smoothed version  of the current day ($z=0$) velocity field derived from the numerical action orbit reconstruction of \citet{2017ApJ...850..207S}.  In brief summary, orbits are followed for 1382 tracers (either groups or individual galaxies) that collectively dominate the mass content associated with luminosity within 38~Mpc $\simeq 2850$~\kms. The tracers are either (or both) important mass constituents or have very accurately known distances (15\% or better uncertainties in the {\it Cosmicflow-3} compendium).  Physically consistent orbits are followed from $z=4$ to  $z=0$.  The orbit reconstruction is embedded in a tidal field at distances greater than 38~Mpc based an the {\it Cosmicflows-2} velocity model of \citet{2014Natur.513...71T}.

There can be no hope of untangling the complexity of orbits after shell crossing, so orbits describe the centers of mass of the ensemble of a group today.  The initial distribution at $z=4$ of our sample is well dispersed, but by today the entities have become concentrated in the current observed structure.  For practical reasons, the density of tracers is highest nearby.  Consequently, the model is most robust within $\sim 15$~Mpc and is poorly sampled in the voids.

The orbits of our Galaxy and Andromeda (M31) are followed separately in the numerical action model of \citet{2017ApJ...850..207S}.  These galaxies are each the central host of distinct collapsed halos. We \citep{2015AJ....149...54T, 2017ApJ...843...16K} equate groups with collapsed halos; the term "Local Group" is a misnomer.  The Galactic Standard of Rest is the natural coordinate system for studies of the orbital history of the Milky Way.

The model distance$-$velocity curve (blue) is derived from interpolation of the {\bf model} velocities on a grid of distances with grid intervals 0.2 Mpc at the origin increasing to 1 Mpc intervals at 38 Mpc, the outer limit of the model.  The value at a grid point is averaged over the four nearest mass points weighted by the inverse square of separation.  Pause the cursor over the blue curve for distance and velocity information at a chosen location.  Pause the cursor over individual entries for information on data points.

\subsection{Linear Model Extending to 200 Mpc}

This second calculator is based on the three-dimensional velocity and density fields of \citet{2019MNRAS.tmp..130G}.  The likelihood model seeks consistency between velocities and the matter distribution in the linear regime of a direct relationship between the gradient of the velocity field and densities.  The procedure is an extension of that by \citet{2016MNRAS.457..172L}.  Care is taken to negate Malmquist bias \citep{1995PhR...261..271S} and the asymmetry in velocity errors in translations to distance from the logarithmic modulus.  Multiple constrained realizations are averaged following \citet{1991ApJ...380L...5H}.

{\it Cosmicflows-3} provides distance and velocity measures.  The 17,647 individual galaxy entries are collected into 11,501 entities through grouping, with two or more galaxies with distance estimates in 1704 groupings.\footnote{The exact numbers of individual and grouped entities change as minor corrections are made.  See http://edd.ifa.hawaii.edu for updated catalogs.}   Linkages are established between galaxies with distance estimates and a 2MASS group catalog \citep{2015AJ....149..171T}.  A group is assigned the weighted average distance of the available measures and the averaged velocity of all known members.   As a consequence, although distances to individual galaxies can have large uncertainties, there are a multitude of groups that are relatively well constrained. The dynamically dominant groups with distance measures are identified by larger symbols in the displays of the linear model in plots analogous to Figure~\ref{DVlin}.\footnote{Beyond 10,000~\kms $\sim 130$~Mpc the group mass assignments from \citet{2015AJ....149..171T} become unreliable because correction factors for missing components become very large.}

The present  \citet{2019MNRAS.tmp..130G} model is constructed on a relatively coarse grid at 6.25~Mpc spacings.  There remains an issue of limited observations at latitudes within $15^{\circ}$ of the plane of our Galaxy.  Nearby, within $\sim 8,000$~\kms, our filtered reconstruction is relatively successful across this zone.  However, at greater distances the gap of the zone of obscuration becomes too large for a good reconstruction.

The Local Sheet (or alternative Local Group) reference frame is preferred in this model because of the cancellation of the Milky Way and M31 motions toward each other.  Averaged velocities are appropriate in the linear regime.  Nearby galaxies have modest dispersions in the Local Sheet frame \citep{2002A&A...389..812K}.  All nearby galaxies share similar large peculiar motions in the frame of the cosmic microwave background.

\section{Practical Example}

As an example of the service that the Distance $-$Velocity calculator provides, consider an effort to determine the Hubble Constant from a measurement of the distance to the gravitational wave event GW170817 that occurred in the galaxy NGC~4993
\citep{2017Natur.551...85A}.  By happenstance, the distance, $d\sim 40$~Mpc, is at the extremity of coverage provided by the non-linear model (although that model is embedded within the 100~Mpc scale tidal field deduced from a linear analysis of {\it Cosmicflows-2}).  It comfortably lies within the domain of coverage of the new linear model.

What is the appropriate velocity to accompany the distance measurement for an estimate of the Hubble Constant?  It is important to understand the environment of the target on a range of scales.  Most immediately, does the target lie in a group?  In the case of NGC~4993, yes, this galaxy lies in a group with the dominant member NGC~4970.  The group is identified as HDC~751 \citep{2007ApJ...655..790C}, 2M++~1294 \citep{2011MNRAS.416.2840L}, Nest~100214 \citep{2015AJ....149..171T}, and PGC1~45466 \citep{2017ApJ...843...16K}.  In the latter catalog, there are 22 galaxies linked with the group, with $<V_{helio}> = 2995\pm25$~\kms.  In alternate reference frames of interest, $<V_{gsr}> = 2851$~\kms, $<V_{ls}> = 2783$~\kms, and $<V_{cmb}> = 3308$~\kms.

\begin{figure*}[]
\centering
\includegraphics[width=0.70\linewidth]{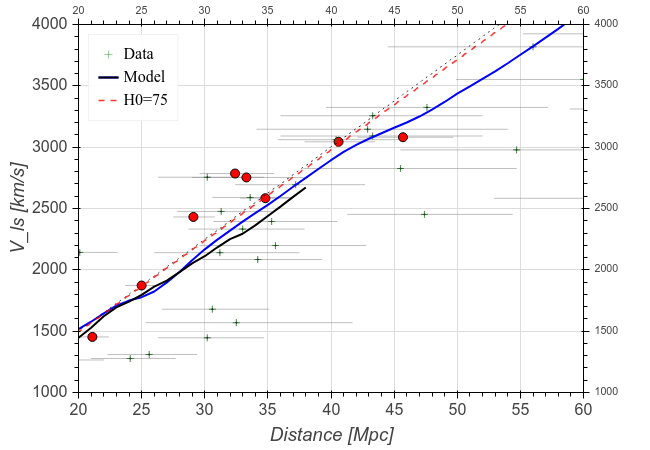}
\caption{
A zoomed extraction from the Distance-Velocity calculator in the direction of the galaxy NGC~4993.  The result from the non-linear calculator terminating at 38~Mpc is overlaid as a black curve on the result from the linear calculator.  Departures from the Hubble expectation with the linear model are $\sim -100$~\kms\ nearward of $\sim 42$~Mpc, increasing in amplitude to $\sim -300$~\kms\ longward of that point.
}
\label{n4993}
\end{figure*}

On a larger scale, this PGC1~45466 = NGC~4970 group lies roughly along the line of sight toward the Great Attractor \citep{1987ApJ...313L..37D}, $18^{\circ}$ ($\sim 13$~Mpc) removed from the dominant Centaurus Cluster ($d=40.5\pm1.6$~Mpc, $V_{ls}=3285$~\kms). There is a clear infall pattern in the line of sight toward the Centaurus Cluster, with objects to the immediate foreground falling away from us toward the cluster and object behind manifesting backside infall toward us.  This effect extends in a weakened fashion to the NGC~4993 line of sight.  At a distance of around 40~Mpc, there is ambiguity whether the NGC~4970 group with NGC~4993 is front-side falling away or back-side falling forward with respect to the overdensity around the Centaurus Cluster.  Figure~\ref{n4993} is extracted from the DV Calculator in the NGC~4993 direction, with the non-linear solution superimposed on the linear solution.  There is close agreement between the two at distances less than 30~Mpc. At greater distances the non-linear model runs about $100$~\kms\ below the linear model but the non-linear model is poorly constrained at its edge of application.  Considering the linear model, within 10~Mpc foreground of $\sim 40$~Mpc observed velocities are running about 100~\kms\ below the H$_0=75$ fiducial line while by 50 Mpc the observed velocities drop to about 300~\kms\ below the fiducial line.  The consequence is an ambiguity due to this Centaurus Cluster overdenity infall pattern. The amplitude of the wave is 200~\kms\ in the linear regime (and greater if non-linear effects would be taken into account).

It is most common to derive estimates of the Hubble Constant in the reference frame of the cosmic microwave background.  However nearby galaxies are participating in a coherent flow, a coherence that extends to include the NGC~4970 group with NGC~4993 as a member.  A correction for the flow is required if working in the cosmic microwave background frame that is closely equivalent to working without a correction in the Local Sheet (Local Group) frame.  This comment is a cautionary reminder of the jeopardy of evaluating the Hubble Constant locally.  Should we live in a large scale under- or over-density, it would be necessary to make measurements beyond this feature to be comfortable in the cosmic microwave background frame.

Finally, It is to be appreciated that the distances in this DV calculator are based on a specific zero-point calibration that may or may not pass the test of time.  The H$_0=75$~\kmsMpc\ fiducial line is consistent with the current {\it Cosmicflows-3} data compilation.  A user of the DV calculator may have a distance that is implicitly based on a different zero-point scale.  However, as discussed in Section 2, peculiar velocities are decoupled from the Hubble Constant.  A user can consider the DV calculator distance scale to be elastic, with the fiducial Hubble line flexing accordingly, but the peculiar velocity amplitudes with respect to the fiducial line will be unchanged at a given observed velocity. 

In review of the case for NGC~4993, as a best effort at accounting for deviations from cosmic expansion, an estimate can be made of the value of the Hubble Constant, H$_0 = (V_{obs}-V_{pec})/d$ taking $V_{obs}=2783$~\kms, the value for the NGC~4970 group, and $V_{pec}$ somewhere between $-100$ and $-300$ depending on the user's preferred distance and whether that places the target foreground or background to the Centaurus Cluster overdensity.

\section{Future Augmentations}

{\it Cosmicflows} is a continuously evolving program.  It is anticipated that the Distance$-$Velocity calculators will be updated at intervals.  The resolution can be improved with greater computational effort.  In combination, the quasi-linear methodology of \citet{2018NatAs...2..680H} can be exploited.  There is the intention of extending the numerical action orbit reconstruction model of \citet{2017ApJ...850..207S} to 100 Mpc.  {\it Cosmicflows-4} is on the horizon, with constraints to be provided by $\sim 30,000$ distances.

The latest models will be made available at the Extragalactic Distance Database.\footnote{http://edd.ifa.hawaii.edu}

\bigskip

\noindent
{\bf Acknowledgements}

Many people have contributed directly or indirectly to this substantial undertaking.  Special thanks to Gagandeep Anand, Igor Karachentsev, Dmitry Makarov, Lidia Makarova, Don Neill, Luca Rizzi, Mark Seibert, Kartik Sheth, and Po-Feng Wu.  This catalog could hardly have been assembled without the resources of NED, the {\it NASA/IPAC Extragalactic Database}, and the Lyon extragalactic database {\it HyperLeda}. 
Financial support for the Cosmicflows program has been provided by the US National Science Foundation award AST09-08846, an award from the Jet Propulsion Lab for observations with {\it Spitzer Space Telescope}, and NASA award NNX12AE70G for analysis of data from the {\it Wide-field Infrared Survey Explorer}.  Additional support has been provided by the Lyon Institute of Origins under grant ANR-10-LABX-66 and the CNRS under PICS-06233 and the Israel Science Foundation grant ISF 1358/18.
  
\bibliography{paper}
\bibliographystyle{aasjournal}

\end{document}